\def\fs{\footnotesize}

\def\al{\alpha}
\def\bt{\beta}
\def\ga{\gamma}
\def\Ga{\Gamma}
\def\de{\delta}

\def\ka{\kappa}
\def\la{\lambda}
\def\La{\Lambda}

\def\si{\sigma}

\def\om{\omega}

\def\pt{\partial}
\def\na{\nabla}

\def\pib{\bar{\pi}}
\def\Lb{\bar{L}}
\def\pb{\bar{p}}

\newtheorem{prop1}{Proposition}[section]
\newtheorem{prop2}[prop1]{Proposition}
\newtheorem{prop3}[prop1]{Proposition}
\newtheorem{prop4}[prop1]{Proposition}
\newtheorem{prop5}[prop1]{Proposition}
\newtheorem{prop6}[prop1]{Proposition}

\documentstyle[11pt]{article}
\begin{document}
\title{UNIVERSALITY OF EINSTEIN EQUATIONS \\FOR THE RICCI SQUARED LAGRANGIANS}

\author{
Andrzej BOROWIEC\thanks{On leave  from the
Institute of Theoretical Physics, University of Wroc{\l}aw,
pl. Maksa Borna 9, 50-204  WROC{\L}AW (POLAND).
}\\
 Marco FERRARIS\\
Mauro FRANCAVIGLIA\\
Igor VOLOVICH\thanks
{Permanent address: Steklov Mathematical
Institute, Russian Academy of Sciences,
Vavilov St. 42, GSP--1, 117966
MOSCOW (RUSSIA).}\\
{\fs Istituto di Fisica Matematica ``J.--L. Lagrange''}\\
{\fs Universit\`a di Torino}\\ {\fs Via C. Alberto 10, 10123 TORINO (ITALY)}
}
\maketitle

\begin{abstract}
It has been recently shown that, in the first order (Palatini) formalism,
there is universality of Einstein equations
and Komar energy--momentum complex, in the sense that for a generic
nonlinear Lagrangian depending only on the scalar curvature of a metric and
a torsionless connection one always gets Einstein equations and Komar's
expression for the energy--momentum complex.
In this paper a similar analysis (also in the framework of the first order
formalism) is performed for all nonlinear Lagrangians depending
on the (symmetrized) Ricci square invariant. The main result is that the
universality of Einstein equations and Komar energy--momentum
complex also extends to this case (modulo a conformal transformation
of the metric).
\end{abstract}
\newpage
\section{Introduction}
\ \ \
\indent It has been recently shown \cite{NC,FFV1} that Einstein equations are
``universal'', in the sense that for a generic nonlinear Lagrangian  $f(R)$
depending only on the scalar curvature $R$ of a metric and a (torsionless)
connection one always gets Einstein equations
if one relies on  the first order  (`` Palatini '')  formalism.
The importance of considering
nonlinear gravitational Lagrangians is particularly stressed  by
the fact that they provide a simple but general approach to  governing
topology \cite{FFV2}; applications to string theory have also been
considered \cite{Vol}; for a review see \cite{MF1,MF2}. Locally, in fact,
one gets standard Einstein
equations from a nonlinear Lagrangian, but there are global topological
effects which are described by nonlinear Lagrangians and which are absent
in the linear case. It is therefore important to investigate the
universality property not only at the level of the equations of motion, but
also for the energy of the gravitational field. This problem was addressed by
us in a previous paper \cite{BFFV}, where it was shown that  universality holds
also for the energy--momentum complex. We calculated in the first order
formalism the covariant energy--density flow for a nonlinear Lagrangian $f(R)$
and we showed that, in a generic case, it equals in fact the Komar expression
already known in the purely metric formalism for the linear Hilbert Lagrangian.

In this paper we shall extend our discussion to the case of Lagrangians with
an arbitrary dependence on the square of the symmetric Ricci tensor constructed
out of a metric and a (torsionless) connection. We shall first suitably
extend the universality theorems of \cite{NC,FFV1} concerning field equations,
by showing that also in this case field equations generically reduce to Einstein
equations (modulo a conformal change of metric);
also the non--generic cases will be
shortly considered (as in \cite{NC,FFV1}). Subsequently we shall calculate
the energy--momentum complex for this family of Lagrangians and we shall
show that it generically reduces to the standard Komar expression, provided
the metric undergoes a conformal rescaling.
In view of further investigations, we shall finally provide some calculations
concerning functions of the square of the full Ricci tensor.

Our results can be relevant also for quantum gravity.
As is well known, in fact, to remove divergences in quantum gravity  one has to add
counterterms to the Lagrangian which depend not only on the scalar curvature but also
on the Ricci and the Riemann tensors invariants. It follows from our results
that "on-shell", in the first order formalism, counterterms depending
on the scalar curvature and the Ricci square invariant generically do not change the
semiclassical limit, since  we still have the standard Einstein equations.

The main result of the present paper is the following:\medskip\\
\noindent {\bf Theorem}\ {\em
Let $S\equiv g^{\mu\nu}g^{\al\bt}R_{(\mu\al)}(\Ga)R_{(\nu\bt)}(\Ga)$ be the
(symmetrized) Ricci square invariant, build out of a metric $g$ and a symmetric
connection $\Ga$. For the action
$$
A(g, \Ga)=\int f(S) \sqrt{g}\,dx  $$
where $f$ is a generic analytic function of one real variable, the Euler--Lagrange
equations (in the first order formalism) lead always to the Einstein equations
$$
R_{\mu\nu}(h) = \ga h_{\mu\nu} $$
for a new metric $h$ which is defined as
$h_{\mu\nu}=\ga^{-1}\,R_{(\mu\nu)}(\Ga)$ -- with a constant $\ga\neq 0$.\\
Moreover, a superpotential for the energy--momentum flow
along any vectorfield $\xi$ is given by the Komar expression
$$
\sqrt{h}\ \nabla^{[ \mu}\xi^{\nu ]}$$}\\
This result will follow from Propositions 2.1, 2.3 and 3.2 below.

It turns out therefore that the action $A(g, \Ga)$ leads to two metrics: besides the
initial metric $g$ one gets a new Einstein metric $h$,	related to $g$ by the
algebraic equation $$
(g^{-1} h)^2=\pm\  I$$
This condition provides on spacetime some
additional structures, namely: a Riemannian almost--product structure or an
almost--complex structure with a Norden metric. These aspects will be studied in more
detail in our forthcoming paper \cite{BFFVA}.

\section{ Field Equations}
\ \ \
\indent As is well known  gravitational Lagrangians which are nonlinear in the scalar
curvature of a metric give rise to higher derivatives or to
the appearance of additional matter fields \cite{Stelle,MFF}.
This strongly depends on having taken
a metric as  basic variable; moreover the equations ensuing
from such Lagrangians show an explicit dependence on
the Lagrangian itself.
It was	shown in \cite{FFV1} that, in contrast,  working in the first
order (Palatini) formalism with independent variations with respect
to a metric and a symmetric  connection, then, for a large
class of Lagrangians of the form $f(R)\,\sqrt g$,
where $R$ is the scalar curvature, the equations obtained are almost independent
on the	Lagrangian, the only dependence being in fact encoded into constants
(cosmological and Newton's ones). In this sense the equations obtained
are ``universal''.
These are in fact Einstein equations in generic cases, while in degenerate
situations and in dimension $n=2$ one gets either equations which express the
constancy of the scalar curvature  or conformally invariant equations.

In this paper we shall consider, still in the first order (Palatini)
formalism, the family of actions
$$
A(\Gamma, g)=\int_M f(S)\sqrt g \ dx \eqno	    (2.1)
$$
\noindent where: $M$ is a n--dimensional (connected) manifold  endowed with a metric
$g_{\mu \nu}$ of an arbitrary signature and a torsionless
(i.e., symmetric) connection $\Ga_{\mu \nu}
^{\sigma}$; the Lagrangian density is $L_{f}=f(S)\sqrt g$ ,
where $f$ is a given function
of one real variable, which we assume to be  analytic;
the scalar $S$ is the symmetric part of a Ricci square--invariant,
considered as a first order scalar concomitant of a metric and the
(torsionless) connection, i.e.:
$$
S \equiv  S(g, \Ga)= S^{\mu \nu}S_{\mu \nu} =
g^{\mu \al}g^{\nu \bt}S_{\al \bt}S_{\mu \nu}	\eqno (2.2)
$$
where $S_{\mu \nu}\equiv S_{\mu\nu}(\Ga)=R_{(\mu \nu)}(\Ga)$ is the symmetric part
of the Ricci tensor. We adopt the standard  notation for the Riemann and Ricci
tensors of $\Ga$ :
$$
R_{\mu \nu \sigma}^{\lambda}(\Ga) \equiv
R_{\mu \nu \sigma}^{\lambda}=\partial_{\nu} \Gamma_{\mu
\sigma}^{\lambda} - \partial_{\sigma}\Gamma_{\mu
 \nu}^{\lambda} + \Gamma_{\alpha\nu}^{\lambda} \Gamma_{\mu
\sigma}^{\alpha} - \Gamma_{\alpha \sigma}^{\lambda}
\Gamma_{\mu \nu}^{\alpha}
$$
$$
R_{\mu \sigma} = R_{\mu \sigma}(\Ga)=R_{\mu \nu \sigma}^{\nu} \ ,\ \
(\alpha, \mu, \nu,...=1,...,n)
$$
and simply write $\sqrt{g}$ instead of $|\det g_{\mu\nu}|^{1/2}$.
Since the Lagrangian $L_f$ does not depend on derivatives of the metric field,
the only momenta are those conjugated to the connection coefficients:
$$
p^{\mu \nu \ka}_{\la} \equiv \frac{\pt L_f}{\pt \Ga_{\mu \nu, \ka}^{\la}}
= \pi^{\al \bt} \frac{\pt S_{\al \bt}}{\pt \Ga_{\mu \nu, \ka}^{\la}}=
 2 (\pi^{\mu \nu}\de^\ka_\la - \pi^{\ka (\mu}\de^{\nu)}_\la) \eqno (2.3)
$$
where $\Ga^\la_{\mu \nu, \ka}= \pt_\ka \Ga^\la_{\mu \nu}$ and
$$
\pi^{\mu \nu} = 2 \sqrt{g}\,f^{\prime}(S)\,S^{\mu \nu} \eqno (2.4)
$$
(recall that $S^{\mu\nu}\equiv g^{\mu\al}g^{\nu\bt}S_{\al\bt}$ but $g^{\mu\nu}$ is
the inverse of $g_{\mu\nu}$).

The Euler--Lagrange equations for the action (2.1) with respect to
independent variations of
$g$ and $\Gamma$ can be written in the following form
$$
f^{\prime} (S)g^{\al \bt}S_{\mu \al}S_{\nu \bt}-{1\over 4}f(S)g_{\mu
\nu}=0 \eqno (2.5)
$$
$$
\na_\al \pi^{\mu \nu} \equiv 2\nabla_{\alpha}(f^{\prime} (S)\sqrt g \
S^{\mu \nu})=0	\eqno (2.6)
$$
\noindent where $ \nabla_{\alpha}$ is the covariant derivative with respect
to $\Gamma$ and we assume $n\geq 2$.
In fact, variation of the action (2.1) with respect to $\Gamma$
gives the equation:
$$
\nabla_{\alpha}\pi^{\mu\nu} -
\nabla_{\rho}\ \pi^{\rho (\mu}\delta^{\nu )}_\alpha =0
$$
\noindent which, due to the symmetry of $\pi^{\mu \nu}$, in any dimension
$n \geq 2$ reduces to (2.6) by taking a trace (see Appendix I).

The properties of the system of equations (2.5)-(2.6) are governed by
the following equation for the scalar $S$:
$$
f^{\prime} (S)S-{n\over4}f(S)=0 \eqno (2.7)
$$
\noindent which is obtained by transvecting (2.5) with $g^{\mu \nu}$.
Following the lines of theorem 1 of \cite{FFV1} we shall then distinguish
the following three mutually exclusive cases: (i) eq. (2.7) has no real
solutions; (ii) eq. (2.7) has isolated real solutions;
(iii) eq. (2.7) is identically satisfied. Their discussion proceeds as follows:
\\\medskip

\noindent {\bf (Case 1 -- no real solutions)} If equation (2.7) has no real solution, then
also the system (2.5)--(2.6) has no real solution.
\\\medskip

\noindent {\bf (Case 2 -- isolated real solutions)} Let us now suppose
that eq. (2.7) is not identically satisfied and has at
least one real solution. In this case, since analytic functions can have
at most a countable set of zeroes on the real line, eq. (2.7)
can have no more than a countable set of solutions $S=c_i$ \ $(i=1,2,\ldots)$,
where $c_i$ are constants.
Consider then any solution
$$
S=c_i \eqno (2.8)
$$
\noindent of eq. (2.7). We have now two possibilities, depending on
the value of the first derivative $f^{\prime}(c_i)$ at the point $S = c_i$.
\\\medskip

\noindent {\bf (Subcase 2.1 -- $f^{\prime}(c_i) \neq 0$)} Let us assume that
$f^{\prime}(c_i) \neq 0$. Then eq. (2.6) takes the form
$$
\nabla_{\alpha}(\sqrt g \ S^{\mu \nu}(\Ga))=0 \eqno (2.9)
$$
\noindent while equation (2.5) reduces to
$$
g^{\al \bt}S_{\mu \al}(\Ga)S_{\nu \bt}(\Ga)=\Lambda(c_i) g_{\mu \nu}
\eqno (2.10)
$$
\noindent where a constant $\Lambda = \Lambda (c_i)$ arises
according to:
$$
\Lambda = \Lambda(c_i) =f(c_i)/4f^{\prime}(c_i)=c_i/n  \eqno (2.11)
$$
\noindent We shall discuss separately the subcases $n>2$ and $n=2$.
\\\medskip

\noindent {\bf (Subcase 2.1.1 -- $f^{\prime}(c_i) \neq 0$ and $n>2$)}
As it is known,
for $n>2$ and any metric $h_{\mu \nu}$, the general solution of the equation
$$
\na^{\Ga}_{\al}(\sqrt{\det h}\,h^{\mu \nu})=0 \eqno (2.12)
$$
is the Levi--Civita connection $\Ga=\Ga_{LC}(h)$ (here $h^{\mu\nu}$ is the
inverse of $h_{\mu\nu}$):
$$
\Ga^\si_{\mu \nu}(h)= {1\over2}h^{\si \al}(\pt_\mu h_{\nu \al} +
\pt_\nu h_{\mu \al} - \pt_\al h_{\mu \nu})
$$
The Ricci tensor $R_{\mu \nu}(\Ga)$ is automatically symmetric and in fact
identical to the Ricci tensor $R_{\mu \nu}(h)$ of the metric $h$ itself.
Hence the following holds true for the present subcase:
\begin{prop1}
Assume $\La\neq0$ and let us take $\ga \neq 0$ and $n > 2$. Then:
\begin{itemize}
\item{(i)}  If $h_{\mu\nu}$ is a metric satisfying the Einstein equations
$$
R_{\mu\nu}(h) =\ga h_{\mu\nu} \
\eqno (2.13)
$$
and  a metric $g_{\mu \nu}$ is an arbitrary solution of the algebraic equation
$$
g^{\al \bt}g^{\nu \rho}h_{\mu \al}h_{\nu \bt}= {\La\over \ga^2}\de^{\rho}_{\mu}
\eqno (2.14)
$$
where $g^{\mu \nu}$  denotes the inverse of $g_{\mu \nu}$,
then the pair $(g,\Ga_{LC}(h))$ satisfies eq.s (2.9)--(2.10).
\item{(ii)} Conversely, if $(g,\Gamma)$ is a solution of eq.s (2.9)--(2.10)
and $\ga \neq 0$, then
$$
h_{\mu \nu}=\ga^{-1}\, S_{\mu \nu}(\Ga)
$$
has to satisfy the algebraic relations (2.14). Moreover $\Ga=\Ga_{LC}(h)$
and one has Einstein equations (2.13).
\end{itemize}
\end{prop1}
\noindent{\it Proof.}\ Assume that $h$ and $g$ are two metrics satisfying
(2.14). In particular this implies
that both metrics are nondegenerate and that
$\det{h^2} =({\La\over \ga^2})^n\,\det{g^2}$, which ensure
proportionality by a constant factor $\det h \sim \det g$. Moreover one finds
$$
(h^{-1})^{\mu\nu}\sim g^{\mu\al}g^{\nu\bt}h_{\al\bt} \eqno (A)$$
which tells us that rising indices of $h$ by means of $g$ produces
a matrix which is proportional to the  inverse of $h$.

Now assume that $h$ is an Einstein metric: $R_{\mu\nu}(h)=\ga\,h_{\mu\nu}$
with $\ga\neq 0$ and take  $\Ga=\Ga_{LC}(h)$. Then
$$S_{\mu\nu}(\Ga)\equiv R_{\mu\nu}(\Ga_{LC})\equiv R_{\mu\nu}(h)=
\ga\,h_{\mu\nu} \eqno (B)$$
Therefore, $\na_\mu(\sqrt h h^{\al\bt})=0$ implies (2.9). The first
claim is thence proved.

Assume conversely that (2.9) and (2.10) are satisfied. Setting
$h_{\mu\nu}=(\ga^{-1}\,S_{\mu\nu})$ it then follows from (2.10)
that $h$ satisfies the algebraic relation (2.14).
Now, using (A),  equation (2.9) can be rewritten in the form
$\na_\al (\sqrt h h^{\mu\nu})=0$. This in turns implies that
$\Ga=\Ga_{LC}(h)$. Therefore, again by (A), $h$ is an Einstein metric and
our claim is proved.\smallskip\hfill Q.E.D.\medskip\\

This extends the results previously found by
Higgs \cite{Hig} for the conformally invariant case $f(S)=S$ in dimension
$n=4$ (see Proposition 2.3), on the basis of an earlier paper by Stephenson
\cite{Step}. The above proposition can be re--phrased as follows.
If we define a new metric $h_{\mu\nu}$ by setting:
$$
h_{\mu\nu} = \ga^{-1}\, S_{\mu\nu}
$$
\noindent with an arbitrary  nonvanishing constant $\ga$,
then eq.s (2.9) and (2.10) imply that $\Gamma$ is the Levi--Civita
connection of the new metric $h$ and eq. (2.10) reduces to
the Einstein eq.s for the metric $h$
$$
R_{\mu\nu}(h) = \ga\, h_{\mu\nu}
$$
\noindent This, in turn, leads to a constant scalar
curvature for the new metric $h$ (in fact, it is $R(h) = n\ga$) and $M$ will
be an Einstein manifold with respect to the new metric $h$.\\ 
\medskip

\noindent{\bf Remark}. Notice that for $\La>0$ eq. (2.14) has an obvious global
solution of the
form $g_{\mu \nu}=e^{-\om}h_{\mu \nu}$ with $\om=1/2\,ln (\La/\ga^2$).
The case $\La \leq 0$ is more complicated and it will be considered in a
forthcoming publication \cite{BFFVA}.\\ \medskip

\noindent{\bf (Subcase 2.1.2 -- $f^{\prime}(c_i) \neq 0$ and $n=2$)}
In the case $n=2$, instead, equation (2.12) allows a further degree
of freedom (related with conformal invariance). It has in fact
the following general solution \cite{NC,FFV2,Des}:
$$
\Gamma_{\mu \nu}^{\sigma}=W_{\mu \nu}^{\sigma}(h,B)=
{1\over2}h^{\sigma \alpha}(\partial_{\mu}h_{\nu \alpha}
+\partial_{\nu}h_{\mu \alpha}-\partial_{\alpha}h_{\mu \nu})
+{1\over 2}(\delta^\si_\nu B_\mu + \delta^\si_\mu B_\nu
	    - h_{\mu\nu}B^\si) \eqno (2.15)
$$
\noindent where $B^\alpha$ is an arbitrary vectorfield and we set
$B_\mu=h_{\mu \al}B^\al$. This is due to
the fact that only in dimension $n=2$ eq. (2.6) cannot be reduced
to $\nabla_{\alpha}h_{\mu \nu}=0$ (since only in two dimensions
it does not imply that the Riemannian volume element of $h$ is covariantly
constant along $\Gamma$). A connection $W(h,B)$
having the form (2.15) is called a {\it Weyl connection} \cite{Wey}.
Using eq. (2.15), from the definition of $S_{\mu\nu}(\Gamma)$ one has:
$$
S_{\mu\nu}(W(h, B))\equiv {1\over 2}(R(h) - D_\alpha B^\alpha)h_{\mu\nu}
\eqno (2.16)
$$
\noindent where $D_\alpha$ denotes the covariant derivative with respect to
the metric $h_{\mu\nu}$, so that eq. (2.13)
reduces to the following scalar equation
$$
R(h) - D_\alpha B^\alpha = 2 \ga
\eqno (2.17)
$$
\noindent Equation (2.17) is the ``universal'' equation
for 2--dimensional space--times \cite{FFV2,Des}; it replaces Einstein equations,
which are the ``universal equations'' in dimension
$n>2$. Equation (2.17) is in fact the equation of constant scalar curvature
for the metric $h$ and the Weyl connection (2.15), because from (2.16)
one has:
$$
{\cal R}(h,B) = h^{\mu\nu} R_{\mu\nu}(W(h,B)) = R(h) - D_\alpha B^\alpha
\eqno (2.18)
$$
\noindent We remark that eq. (2.17) has always infinitely
many local solutions, but it might have no global analytic solution
(depending on the topology of the 2-dimensional manifold $M$). In any case,
the following holds true for the subcase 2.1.2:
\begin{prop2}
Assume $\La\neq0$, $\ga\neq0$ and $n=2$. Then:
\begin{itemize}
\item{(i)}  If $h_{\mu\nu}$ is a metric satisfying equation (2.17)
where $B^\al$ is an arbitrary vectorfield and a metric $g_{\mu \nu}$ is an arbitrary
solution of the algebraic eq. (2.14),
then the pair $(g, W(h, B))$ satisfies eq.s (2.9)--(2.10).
\item{(ii)} Conversely, if $(g,\Gamma)$ is a solution of eq.s (2.9)--(2.10),
then
$$
h_{\mu \nu} = \ga^{-1}\,S_{\mu \nu}(\Ga)
$$
has to satisfy the algebraic relations (2.14). Moreover $\Ga=W(h, B)$
for some vectorfield $B^\al$ and one has eq.s (2.17).
\end{itemize}
\end{prop2}
{\it Proof}\ . The proof goes along the same lines as in Proposition 2.1,
with the only difference that Einstein eq.s are now replaced
by equations (2.17).\\\smallskip\hfill Q.E.D.\smallskip\\

\noindent {\bf (Subcase 2.2 -- $f^{\prime}(c_i) = 0$)}
Suppose now $f^{\prime}(c_i)=0$.
Then eq. (2.7) implies that also $f(c_i)=0$, i.e., $c_i$ is a zero of order at
least two of $f(S)$. In this case, eq.s (2.5)--(2.6) are identically
satisfied and the only relation between $g$ and $\Gamma$ is contained
in the following equation
$$
S(g,\Gamma) = c_i     \eqno (2.19)
$$
\noindent This equation represents a genuine dynamical relation between the
metric and the connection of $M$, although it is not enough to single
out a connection $\Gamma$ for any given metric $g$ (as it happened,
instead, in subcase (2.1) above, where $\Gamma$ turns out to be the
Levi--Civita connection of $h$ and $h$ is algebraically related to $g$).
In fact, defining a tensorfield $\Delta^{\lambda}_{\mu\nu}$ by:
$$
\Delta^{\lambda}_{\mu\nu} \equiv \Delta^{\lambda}_{\mu\nu}(g,\Gamma) =
\Gamma^{\lambda}_{\mu\nu} - \Gamma^{\lambda}_{\mu\nu}(g)
\eqno (2.20)
$$
\noindent equation (2.19) can be turned into a quasi--linear first order PDE for
the unknown $\Delta$, having the term $S(g) - c_i$ as a source. The
space of solutions of this last equation, as functions of the metric
$g$ together with its first derivatives and a number of auxiliary fields,
has a complicated structure. In any case, it is easy to see that this space
contains as a subspace the space of all pairs $(g,\Gamma)$
satisfying eq. (2.10) for $\Lambda = {c_i/ n}$. We also remark that for
$c_i=0$ this equation covers the case $\La=0$ which was excluded from
Proposition 2.1 (see above).\medskip\\

\noindent {\bf (Case 3 -- the conformally invariant case)}
We consider now the case in which eq. (2.7) is
identically satisfied. Under this hypothesis the Lagrangian is
proportional to:
$$
f(S)=S^{n/4} \eqno (2.21)
$$
\noindent
We shall then consider for simplicity the case $S \geq 0$ (analogous results
will be valid for $S \leq 0$ and they may extend across $S=0$ at least
if $n=4k$). We stress that for $n=4$ this is exactly the linear
Lagrangian $f(S)=S$ considered in \cite{Hig,Step}.
\par

In this case equations (2.5) and (2.6) read as follows:
$$
 S^{n-4\over 4}(g^{\al \bt}S_{\mu \al}
     S_{\nu \bt} - {4\over n} S g_{\mu\nu} ) = 0
\eqno (2.22)
$$
$$
\nabla_\alpha (S^{n-4\over 4}
    \sqrt g S^{\mu\nu} ) = 0 \eqno (2.23)
$$
\noindent Notice first of all that, under conformal transformations
$$
\tilde g_{\mu\nu} = e^{\omega}g_{\mu\nu} \  , \   \tilde \Gamma = \Gamma
\eqno (2.24)
$$
\noindent where $\omega$ is an arbitrary function on manifold $M$, one has
\begin{eqnarray*}
\tilde S &=&  e^{-2\omega}S \,\\
\tilde S_{\mu\nu} &=& S_{\mu\nu} \,\\
\tilde S^{n/4} \sqrt{\tilde g} &=& S^{n/4}\sqrt g \,\\
\tilde S^{n-4\over 4} \sqrt{\tilde g}\tilde S^{\mu\nu} &=&
S^{n-4\over 4} \sqrt g S^{\mu\nu} \,
\end{eqnarray*}
\noindent Therefore, the action
$$
A(g,\Gamma) = \int_M S^{n/4} \sqrt g\, dx
$$
\noindent as well as the equations (2.22) and (2.23) are invariant under the
transformation (2.24), i.e., $A(\tilde g,\tilde \Gamma) = A(g,\Gamma)$.

Also in this case we shall consider separately the two
cases $n>2$ and $n=2$.\\\medskip

\noindent {\bf (Subcase 3.1 -- conformally invariant for $n>2$)}
  If $n>2$ there are two possibilities. If $S=0$ everywhere on our manifold
then we have only the equation
$$
S(g,\Gamma) = 0
$$
\noindent and the discussion proceeds as for eq. (2.19) above. When $S\neq0$
there is instead an additional conformal degree of freedom, as noticed earlier
in \cite{Hig} and \cite{Step} (and later exploited in \cite{FE}).
In what follows we restrict ourselves to the case when $S$ is a nowhere vanishing
function. Then, in fact, eq.s (2.22) and (2.23) reduce to the following:
$$
g^{\al \bt}S_{\mu\al} S_{\nu \bt} - {4\over n} S g_{\mu\nu} = 0
\eqno (2.25)
$$
$$
\nabla_\alpha (S^{-1}
    \sqrt{\det{S_{\al\bt}}}\, S^{\mu\nu} ) = 0 \eqno (2.26)
$$
\noindent In order to get (2.26) we have used (2.25) to calculate first the determinant
of $g$ as a function  of $\det{(S_{\al\bt})}$ and next replace it into (2.23).
The following proposition is true:

\begin{prop3}
Assume $S\neq 0$, $\ga\neq0$ and $n > 2$. Then:
\begin{itemize}
\item{(i)}  If $h_{\mu\nu}$ is a metric satisfying the Einsten equations
$$
R_{\mu\nu}(h) =\ga h_{\mu\nu} \
$$
where a metric $g_{\mu \nu}$ is an arbitrary solution of the algebraic equation
$$
g^{\al \bt}g^{\nu \rho}h_{\mu \al}h_{\nu \bt}= {4s\over n\ga^2}
\de^{\rho}_{\mu} \eqno (2.27)
$$
with an arbitrary nowhere vanishing function $s$,
then the pair $(g,\Ga_{LC}(h))$ satisfies eq.s (2.25)--(2.26).
Moreover one has $S(g, \Ga_{LC}(h))=s$.
\item{(ii)} Conversely, if $(g,\Gamma)$ is a solution of eq.s (2.25)--(2.26),
then
$$
h_{\mu \nu}= \ga^{-1}\, S_{\mu \nu}(\Ga)
$$
has to satisfy the algebraic relations (2.27) with $s=S(g, \Ga)$.
Moreover one has $\Ga=\Ga_{LC}(h)$ and $R_{\mu\nu}(h) =\ga h_{\mu\nu}$.
\end{itemize}
\end{prop3}
{\it Proof.}\ Since $S$ is a nowhere vanishing function on a (connected) manifold,
it is strictly positive or strictly negative.
Therefore, it can be written in the form $S=\pm\,e^{2\omega}$. By the
conformal change of metric $g\mapsto e^\omega g$, equation (2.25) transforms
into (2.14)  and equation (2.26) transforms into (2.12). (Notice that after
the conformal transformation $S$ transforms to $1$). Then the same reasoning
as in Proposition 2.1 proves our claim.\medskip\hfill Q.E.D.\medskip\\

\noindent{\bf Remark}. Notice that for $s>0$ eq. (2.27) has an obvious global
solution of the form $g_{\mu \nu}=e^{-\om}h_{\mu \nu}$ with a conformal
factor $2\,\om=ln (4s/ n\ga^2)$.\\\smallskip

We can also restate the
result as follows: if $(g,\Gamma)$ is a solution of eq.s (2.25) and (2.26),
then there exists a scalar field $\psi$ such that the conformally related
metric $\psi g_{\mu\nu}$ satisfies Einstein equations and the connection
$\Gamma$ is the Levi--Civita connection of $\psi g_{\mu\nu}$; moreover,
the scalar curvature of the original metric $g$ and the connection
$\Gamma$ equals $\psi$.
The origin of this extra scalar field $\psi=e^{-\om}$ will be discussed elsewhere
\cite{FMV1} in the framework of Legendre transformation  for metric--affine
theories.\medskip\\

\noindent {\bf (Subcase 3.2 -- conformally invariant and $n=2$)}
If $n=2$ equations (2.18) and (2.19) simplify to
$$
g^{\al \bt} S_{\mu \al} S_{\nu \bt}(\Gamma) - {1\over 2} S(g,\Gamma) g_{\mu\nu} = 0
\eqno	(2.28)
$$
$$
\nabla_\alpha (S^{-{1\over 2}} \sqrt{g} S^{\mu\nu} ) = 0
\eqno(2.29)\\
$$
Then the following holds true:
\begin{prop4}
Assume $S\neq0$, $\ga\neq0$ and $n=2$. Then:
\begin{itemize}
\item{(i)}  If $h_{\mu\nu}$ is a metric satisfying equation (2.17)
where $B^\al$ is an arbitrary vectorfield and a metric $g_{\mu \nu}$ is an
arbitrary  solution of the algebraic eq.(2.27),
with an arbitrary nowhere vanishing function $s \in {\cal F}(M)$,
then the pair $(g, W(h, B))$ satisfies eq.s (2.28)--(2.29).
Moreover $S(g, W(h, B))=s$.
\item{(ii)} Conversely, if $(g,\Gamma)$ is a solution of eq.s (2.28)--(2.29),
then
$$
\ga\,h_{\mu \nu} = S_{\mu \nu}(\Ga)
$$
has to satisfy the algebraic relations (2.27) with $s=S(g, \Ga)$.
Moreover $\Ga=W(h, B)$ for some vectorfield $B^\al$ and one has (2.29).
\end{itemize}
\end{prop4}
{\it Proof}.\  The proof is an obvious combination of those already given for
Propositions 2.2 and 2.3 . \hfill Q.E.D.\medskip\\
\section{Energy--Momentum Complex}%
\ \ \
\indent We are now ready to apply general formulae for the energy--momentum complex
to the Lagrangian density
$$
L=L(g_{\mu \nu}, \Gamma^{\lambda}_{\mu \nu}, \Gamma^{\lambda}_{\mu \nu ,
\alpha})= f(S)\ \sqrt g  \eqno (3.1)
$$
These general formulae are considered in Appendix II and follow closely
the earlier works of ours \cite{FF1,FF2,FF3}; see also \cite{W,WL}. Our
results will extend those of our previous paper \cite{BFFV}.

Our field variables are now $g$ and $\Gamma$, while the only derivatives
entering the Lagrangian are
the first order derivatives of $\Gamma$. Therefore, from (II-1) one has the
following definition of the energy--density flow
$$
E^{\alpha}= p^{\mu \nu \alpha}_{\lambda}{\cal L}_{\xi}\Gamma^{\lambda}
_{\mu \nu} - L\ \xi^{\alpha}	\eqno (3.2)
$$
where $p^{\mu \nu \alpha}_{\lambda}$ are given by (2.3)--(2.4).
By using the decomposition (see II--6):
$$
{\cal L}_{\xi}\Gamma^{\kappa}_{\mu \nu} = -\xi^{\alpha}R^{\kappa}_{
( \mu \nu ) \alpha} + \nabla_{(\mu}\nabla_{\nu)}\xi^{\kappa}
$$
and taking the symmetry of $p^{\mu \nu \lambda}_{\kappa}$ into account, one has
$$
p^{\mu \nu \lambda}_{\kappa}{\cal L}_{\xi}\Gamma^{\kappa}_{\mu \nu}=
- p^{\mu \nu \lambda}_{\kappa}R^{\kappa}_{\mu \nu \alpha}\xi^{\alpha}
+ p^{\mu \nu \lambda}_{\alpha}\nabla_{\mu}\nabla_{\nu}\xi^{\alpha}\ \
$$
where $\na_\mu$ is the covariant derivative with respect to the connection
$\Ga$. From this (see (II--8)) one can easily recognize that:
$$
T^{\lambda \sigma}_{\alpha}=0\ \ \ , \ \ \ \ \ \ T^{\lambda \sigma \rho}_{
\alpha} = p^{\sigma \rho \lambda}_{\alpha}
$$
Substituting these into (II--13) and making use of (2.3)--(2.4) one gets finally
$$
U^{\mu \nu}(\xi)= 2\pi^{\sigma [ \mu}\nabla_{\sigma}\xi^{\nu ]}
+ (\nabla_{\sigma}\pi^{\sigma [\mu})\xi^{\nu]}	   \eqno   (3.3)
$$
The second term of (3.3) vanishes on shell by virtue of (2.6).
Therefore we have proved the following:

\setcounter{prop1}{0}
\begin{prop5}
The energy--density flow (3.2) for the Lagrangian (3.1) can be represented
on shell under the form $E^{\mu}= d_{\nu}U^{\mu \nu}$,	where the
superpotential $U^{\mu \nu}$ is given by:
$$
U^{\mu \nu}= 2\  \pi^{\sigma [ \mu}\ \nabla_{\sigma}\xi^{\nu ]}
\eqno (3.4)
$$
\hfill Q.E.D.
\end{prop5}

Let us now discuss the expression  (3.4) we found for the superpotential. It resembles
very much Komar superpotential, which is known to be a good superpotential
in the purely metric formalism \cite{Komar} (Komar superpotential is in fact
equal also to the Schr\"{o}dinger--M\"{o}ller--Mizkievich superpotential;
for a discussion see, e.g., \cite{Mizk,Traut}).
Our superpotential differs in fact from Komar's one by a factor.
If the dimension $n$ of space--time is larger than $2$
and we are in a generic position (see Section 2), i.e. if equation (2.4) has a
solution $S=c$ such that $c\neq 0$ and $f^{\prime}(c) \neq 0$, then we know
that $\Gamma = \Gamma_{LC}(h)$ for some $h$ given by eq. (2.14). Accordingly,
from (2.4) and (2.14) one has
$$
\pi^{\mu \nu}=\ 2\ga\,({\ga^2 \over \La})^{n\over4} \sqrt{h}\ h^{\mu \nu}
\eqno (3.5)
$$
and (3.4) reduces to:
$$
U^{\mu \nu} =\ 4\ga\,({\ga^2 \over \La})^{n\over4}
\sqrt{h}\ \na^{[ \mu}\xi^{\nu ]} \eqno (3.6)
$$
which is proportional to the Komar expression for $h$.
This corresponds to the subcase 2.1.

At bifurcation points $S=c$, i.e., in the subcase (2.2),
one has instead $f^{\prime}(c) = 0$.
Therefore in this case one has $\pi^{\mu \nu} = 0$
and the energy--density flow as given by (3.4) is identically vanishing.

For the conformally invariant Lagrangian $L = S^{n/4}\, \sqrt{g}$
(which in dimension $n=4$ is just the quadratic Lagrangian
$S\,\sqrt g$) one has finally:
$$
2\,\pi^{\mu \nu} = n\,S(g, \Gamma)^{\frac{n-4}{4}}\ \sqrt{g}\ S^{\mu \nu}
\eqno{}
$$
and one can use a conformal transformation $h=e^\om g$ (for $S \neq 0$) to
reduce the system to Einstein equations (this is in fact the subcase 3.1).
Otherwise, one can decompose the connection $\Gamma$ as in (2.20):
$$
\Gamma^{\rho}_{\mu \nu}(h) = \Ga^{\rho}_{\mu \nu}(g) +
\Delta^{\rho}_{\mu \nu}
$$
where the tensor $\Delta\equiv\Delta (\om)$
is symmetric in its lower indices. The superpotential takes then the form
$$
U^{\mu \nu}= 2\  \pi^{\sigma [ \mu}\,\tilde\na_{\sigma}\xi^{\nu ]}
+ 2\  \pi^{\sigma [ \mu}\ \Delta^{\nu ]}_{\sigma \alpha}\xi^{\alpha}
\eqno (3.7)
$$
where $\tilde{\na}$ is the covariant derivative with respect to the metric $g$.
As we mentioned above, the tensor $\Delta$ can be eliminated by a suitable conformal
transformation, so that the relevant part of the superpotential turns out to
be again proportional, modulo a conformal rescaling of the metric, to the Komar
superpotential of $h$. Therefore one gets
\begin{prop6}
In the cases described by Propositions 2.1 and 2.3 above the superpotential
turns out to be proportional to the Komar superpotential of $h$, {\it i.e.}
$$
U^{\mu\nu}\sim \sqrt h \na^{[\mu}\xi^{\nu]}
$$
\hfill Q.E.D.
\end{prop6}\medskip

The two--dimensional case is particularly important (see \cite{FFV2,Vol,Des}).
In this case $\Gamma$ is a Weyl connection with (see (2.7)):
$$
\Delta^{\rho}_{\mu \nu} = \delta^{\rho}_{( \mu}B_{\nu )} -
 1/2\,h_{\mu \nu}B^{\rho}
$$
and the superpotential reduces to:
$$
U^{\mu \nu} \sim \sqrt{h}\, (D^{[ \mu}\xi^{\nu	]}
 - B^{[ \mu}\xi^{\nu ]}) \ \  \eqno (3.8)
$$
This covers subcases 2.1.2 (see Proposition 2.2) and 3.2 (see Proposition 2.4).

A discussion of the energy--momentum complex for non--linear Lagrangians in
the first order formalism, based on the use of a background connection as
in \cite{FF4}, will be discussed elsewhere \cite{FFB}.

\section{Conclusions}
\ \ \
\indent In this paper it was shown that the universality of Einstein equations and
Komar energy-momentum complex, already found in previous investigations about
the first order formalism, for
nonlinear Lagrangians depending on the scalar curvature,
extends also to Lagrangians depending on the Ricci square invariant.
We have preferred to limit ourselves here to discuss in detail various
important particular cases and emphasize the general picture. It seems to us
that the universality of Einstein equations and Komar energy-momentum
complex for nonlinear Lagrangians depending on the scalar curvature or the Ricci
square invariant is not only interesting and important as a mathematical results
but also in view of furthers applications to concrete problems in gravity
theories (including cosmological models), both at classical and quantum level.
Moreover our results find interesting applications to signature change in spacetime.
We will discuss these matters, as well as the algebraic relations between the two
metrics $g$ and $h$ in forthcoming papers. It will be in particular shown in
\cite{BFFVA} that we obtain, in fact,  new interesting geometrical structures on
manifold, so called almost-product and almost-complex Einsteinian structures.

\section*{Acknowledgments}
\ \ \
\indent Two of us (A.B. and I.V.) gratefully acknowledge
the hospitality of the Institute of Mathematical Physics ``J.--L. Lagrange''
of the University of Torino and
the support of G.N.F.M. of
Italian C.N.R. This work is sponsored by G.N.F.M., M.U.R.S.T. (40\% \ Proj.
``Metodi Geometrici e Probabilistici in Fisica Matematica''); one of us
(A.B.) acknowledges also the support from KBN 2 P302 023 07.

\section*{Appendix I}
\ \  \
\indent In this Appendix we apply the  formalism to the case in which the
scalar $Q$ is equal to the full square of Ricci tensor,
{\it i.e.} when $Q = R^{\mu \nu}\,R_{\mu \nu}$. This is in fact the case studied
by Higgs and Stephenson
\cite{Hig,Step} for the Lagrangian $L=Q\,\sqrt g$ and our result explains
the oversimplification made by Higgs (following Stephenson).
As above, we shall then consider the general class of Lagrangians
$\Lb \equiv \Lb_f = f(Q)\,\sqrt g$, where
$$
Q(g, \Ga)= g^{\mu \al}g^{\nu \bt}R_{\mu \nu}(\Ga)R_{\al \bt}(\Ga)
\eqno (\hbox{I}-1)
$$
is the Ricci square--invariant. Instead of (2.3) we have the following
expression for the momentum
$$
\pb^{\mu \nu \ka}_{\la} \equiv \frac{\pt \Lb}{\pt \Ga_{\mu \nu, \ka}^{\la}}
= \pib^{\al \bt} \frac{\pt R_{\al \bt}}{\pt \Ga_{\mu \nu, \ka}^{\la}}=
 2 (\pib^{\mu \nu}\de^\ka_\la - \pib^{\ka (\mu}\de^{\nu)}_\la) \eqno (\hbox{I}-2)
$$
where
$$
\pib^{\mu \nu} = 2 \sqrt{g}\,f^{\prime}(Q)\,R^{\mu \nu} \eqno (\hbox{I}-3)
$$
and the tensor density $\pib^{\mu \nu}$ is no longer symmetric. (Notice that
$\pb^{\mu \nu \ka}_\la$ maintains instead the symmetry, i.e. $\pb^{[\mu \nu] \ka}_\la =0$,
due to the symmetry of $\Ga$).
The Euler--Lagrange equations  with respect to independent variations of
$g$ and $\Gamma$ can thence be written in the following form
$$
f^{\prime} (Q)g^{\al \bt}(R_{\mu \al}R_{\nu \bt} +
R_{\al \mu}R_{\bt \nu})-{1\over 2}f(Q)g_{\mu \nu}=0 \eqno (\hbox{I}-4)
$$

$$
\nabla_{\alpha}\pib^{(\mu\nu)} -
\nabla_{\rho}\ \pib^{\rho (\mu}\delta^{\nu )}_\alpha =0
 \eqno (\hbox{I}-5)
$$
Accordingly, the ``characteristic equation'' for (I--4)  is given by
$$
4\,f^{\prime}(Q)Q\,-\,n\,f(Q)\,=\,0 \eqno (\hbox{I}-6)
$$
Now, being $\pib^{\mu \nu}$ not symmetric, eq. (I-5) cannot in general be replaced by
$$
\na_\al \pib^{(\mu \nu)}=0 \eqno (\hbox{I}-7)
$$
as it was done in \cite{Hig,Step}. Of course, any solution of
(I--4), (I--7), under the additional assumption $\pt_\al\pib^{[\al\bt]}=0$,
is a solution of (I--4), (I--5), but the general
solution is more complicated (see \cite{FK1}, \cite{FK2} and \cite{Fer} for $n=4$).
The general solution of (I--5) is the following ($n>2$):
$$
\Ga^\al_{\bt\mu} = \Ga^\al_{\bt\mu}(h) + {1\over 2}(\de^\al_\mu A_\bt +
\de^\al_\bt A_\mu -(n-1)A^\al h_{\bt\mu}) \eqno (\hbox{I}-8)$$
where $h_{\al\bt}$ and $A_\mu$ are defined by
$$
\sqrt{h} h^{\al\bt}=\pib^{(\al\bt)}, \ \ \ A_\al=h_{\al\bt}A^\bt$$
and
$$
A^\al= {4\over{(n-1)(n-2)\sqrt{h}}}\pt_\bt\pib^{[\bt\al]}$$

\section*{Appendix II}
\ \ \
\indent In this Appendix we recall some formulae related to the energy--momentum
complex for the first order formalism (see \cite{BFFV,Murphy,Novotny}.

Let us consider on $M$	a set of fields $\phi^{i}$, which we assume
to be fields of geometric objects over $M$,
together with a Lagrangian density $ L(\phi^{i}, \phi^{i}_{\mu}) $  depending
on the first derivatives  $\phi_{\mu}^{i}= \partial_{\mu} \phi^{i}$.
According to \cite{FF1,FF2,FF3}
the canonical energy-density flow $E^{\la}$ is defined by the formula
$$
E^{\lambda} \equiv E^\lambda (\xi)= p^{\lambda}_{i}{\cal L}_{\xi}\phi^{i}
- L \xi^{\lambda} \eqno (\hbox{II}-1)
$$
where $\xi=\xi^{\alpha}\partial_{\alpha}$ is an arbitrary vectorfield
on $M$, ${\cal L}_{\xi}$ is the Lie derivative and the momenta
$p^{\lambda}_{i}$ are defined by
$$
p^{\lambda}_{i}= \frac{\partial L}{\partial \phi^{i}_{\lambda}} \eqno (\hbox{II}-2)
$$
The  ``weak conservation law''
$$
d_{\lambda}E^{\lambda}=0 \eqno (\hbox{II}-3)
$$
holds along solutions of the equations of motion
(here, and in the sequel, $d_{\la}$ denotes ``total'' derivative
with respect to $x^{\la}$).
We assume also that the
Lagrangian is  natural, in the sense that the lift of any
diffeomorphism of $M$ is a symmetry for $L$. This means that one has
$$
{\cal L}_{\xi}L \equiv \partial_{\alpha}(\xi^{\alpha}L) \ = \
\frac{\partial L}{\partial \phi^{i}}{\cal L}_{\xi}
\phi^{i}+ \frac{\partial  L}{\partial \phi^{i}_{\alpha}}{\cal L}_
{\xi}\phi^{i}_{\alpha} \eqno (\hbox{II}-4)
$$
As a consequence, the first variational formula can be written as follows:
$$
\partial_{\mu}E^{\mu} = -\ \frac{\delta L}{\delta \phi^i}
{\cal L}_{\xi}\phi^i   \eqno (\hbox{II}-5)
$$
where $\delta L / \delta \phi^i$ denotes the variational derivative of $L$.
Expanding the Lie
derivative ${\cal L}_{\xi}\phi^i$ as a linear combination of $\xi$ and
its covariant derivatives with respect to an arbitrary symmetric linear
connection on $M$, i.e.
$$
{\cal L}_{\xi}\phi^i = \Phi^{i}_{\alpha}\xi^{\alpha} +
\Phi^{i \rho}_{\alpha}\nabla_{\rho}\xi^{\alpha} +
\Phi^{i \rho \sigma}_{\alpha}\nabla_{\rho}\nabla_{\sigma}\xi^{\alpha}
\eqno (\hbox{II}-6)
$$
the first variational formula can be rewritten in the form
$$
d_{\mu} E^{\mu} = W  \eqno (\hbox{II}-7)
$$
where
$$
E^{\mu} \equiv E^{\mu} (\xi)= T^{\mu}_{\alpha}\xi^{\alpha}+
T^{\mu \rho}_{\alpha}\nabla_{\rho}\xi^{\alpha} +
T^{\mu \rho \sigma}_{\alpha}\nabla_{\rho}\nabla_{\sigma}\xi^{\alpha}
\eqno (\hbox{II}-8)
$$
\newline
$$
W  \equiv  W (\xi) = W_{\alpha}\xi^{\alpha}+
W^{\rho}_{\alpha}\nabla_{\rho}\xi^{\alpha} +
W^{\rho \sigma}_{\alpha}\nabla_{\rho}\nabla_{\sigma}\xi^{\alpha}
\eqno (\hbox{II}-9)
$$
The coefficients of $W$ vanish on shell  because of Euler--Lagrange
equations, so that (II-3) holds on shell; moreover,
they satisfy so called ``generalized Bianchi identities'':
$$
W_{\mu} - \nabla_{\nu}W^{\nu}_{\mu} + \nabla_{\nu}\nabla_{\rho}W^{\nu \rho}
_{\mu} = 0  \eqno (\hbox{II}-10)
$$

Notice that in the formulae (II-6),  (II-8) and (II-10), due to the
uniqueness of the decomposition, one assumes the
symmetry conditions for the highest coefficients, i.e.:
$$
\Phi^{i [\rho \sigma]}_{\alpha}=0,\  T^{\mu [ \rho \sigma ]}_{\alpha} =0,
\ W ^{ [ \rho \sigma ]}_{\alpha} = 0
$$
Therefore, the energy--density flow admits a representation
$$
E^{\mu} = \tilde{E}^{\mu} \ +\ d_{\rho}U^{\mu \rho} \eqno (\hbox{II}-11)
$$
where $U^{\mu \rho} \equiv U^{\mu \rho}(\xi)$ is a skew--symmetric tensor
density (called a ``superpotential'') and $\tilde{E}^{\mu} \equiv \tilde{E}^{\mu}
(\xi)$ is the ``reduced'' energy-momentum flow, which vanishes on shell
(see \cite{FF3}). The following formulae hold true :
$$
\tilde{E}^{\mu} = (W^{\mu}_{\alpha} - \nabla_{\rho}W^{\rho \mu}_{\alpha
}) \xi^{\alpha} + W^{\mu \rho}_{\alpha}\nabla_{\rho}\xi^{\alpha} \eqno (\hbox{II}-12)
$$
\newline
$$
U^{\mu \rho}=(T^{[ \mu \rho ]}_{\alpha} + \nabla_{\nu} \tilde{T}^{\nu [
\mu \rho ]}_{\alpha})\xi^{\alpha} + \tilde{T}^{\mu \rho \nu}_{\alpha
}\nabla_{\nu}\xi^{\alpha}
 \eqno (\hbox{II}-13)
$$
where, $\tilde{T}^{\mu \rho \nu}_{\alpha}= (4/3)T^{[ \mu \rho ] \nu}_{
\alpha}$ (recall that $T^{\mu [ \rho \nu ]}_{\alpha}=0$).


\begin{thebibliography}{}
\bibitem{NC}
M. Ferraris, M. Francaviglia and I. Volovich,
Nuovo Cimento,\  {\bf 108B}\, 1993, p. 1313

\bibitem{FFV1}
M. Ferraris, M. Francaviglia and I. Volovich,
Class. Quant. Grav. {\bf 11}\, 1994, p. 1505

\bibitem{FFV2}
M. Ferraris, M. Francaviglia and I. Volovich, {\it A Model of
Topological Affine Gravity in Two Dimensions and Topology Control},
preprint {\bf TO--JLL--P 2/93} (Torino, June 1993);\ gr-qc/9302037

\bibitem{Vol}
I. Volovich, Mod. Phys. Lett. {\bf A8}\, 1993, p. 1827

\bibitem{MF1}
M. Francaviglia, {\it A New Action for Einstein Equations and
Two-Dimensional Gravity}, Proceedings of the Second International Sakharov Conference
in Physics, Moscow, May 1996 (in print)

\bibitem{MF2}
M. Francaviglia, {\it First Order Non-Linear Action for General Relativity and
Two-Dimensional Gravity}, General Relativity and Gravitational Physics -- Proceedings
of the 12th Italian Conference, Ed. E. Coccia at al., Rome, September 1996 (in print)

\bibitem{BFFV}
A. Borowiec, M. Ferraris, M. Francaviglia and I. Volovich,
Gen. Rel. Grav. {\bf 26}(7)\, 1994, p. 637

\bibitem{BFFVA}
A. Borowiec, M. Ferraris, M. Francaviglia and I. Volovich,
{\it Almost Complex and Almost Product	Einstein Manifolds from a
Variational Principle} (in preparation)

\bibitem{Stelle}
K.A. Stelle, Gen. Rel. Grav. {\bf 5}\, 1978, p. 353

\bibitem{MFF}
G. Magnano, M. Ferraris and M. Francaviglia, Gen. Rel. Grav.
{\bf 19}(5)\, 1987, p. 465

\bibitem{Hig}
P. W. Higgs, Nuovo Cimento {\bf 11}(6)\, 1959, p. 816

\bibitem{Step}
G. Stephenson, Nuovo Cimento {\bf 9}\, 1958, p. 263

\bibitem{Des}
S. Deser, {\it Inequivalence of First and Second Order Formulations
in D=2 Gravity Models}, gr-qc/9512022

\bibitem{Wey}
H. Weyl, {\it Raum--Zeit--Materie}, {\bf 4}th ed., Springer, (Berlin, 1923)

\bibitem{FE}
M. Ferraris, in: Atti del VI Conv. Naz. di Rel. Gen. e Fis. della Grav.
(Florence, 1984), pp. 127--136; ed. M. Modugno; Tecnoprint (Bologna,
1986)

\bibitem{FMV1}
M. Francaviglia, G. Magnano and I. Volovich  (in preparation)

\bibitem{FF1}
M. Ferraris and M. Francaviglia,
J. Math. Phys. {\bf 26}(6)\, 1985, p. 1243

\bibitem{FF2}
M. Ferraris and M. Francaviglia,
Class. Q. Grav.  {\bf 9}, Supplement 1992, p. S79

\bibitem{FF3}
M. Ferraris, M. Francaviglia and O. Robutti,
in:   {\it G\'eom\'etrie et
Physique (Proceedings  Journe\`ees Relativistes  de Marseille 1985)};
Y. Choquet-Bruhat, B. Coll, R. Kerner and A. Lichnerowicz  eds.;
Travaux en Cours, Hermann (Paris, 1987), pp.112-125

\bibitem{W}
R. M. Wald,
J. Math. Phys. {\bf 31}\, 1990, p. 2378; Phys. Rev. {\bf D 50}\, 1994,
p. 846

\bibitem{WL}
J. Lee and R. M. Wald,
J. Math. Phys. {\bf 31}\, 1990, p. 725

\bibitem{Komar}
A. Komar, Phys. Rev. {\bf 113}(3)\, 1959, p. 934

\bibitem{Mizk}
N.V. Mizkievich, A.P. Efremov and A.I. Nesterov, {\it Fields Dynamic in
General Relativity}, Moscow (Nauka, 1989; in Russian)

\bibitem{Traut}
A. Trautman, {\it Conservation Laws in General Relativity}, in:
{\it Gravitation, An Introduction to Current Research}; ed. L. Witten,
Wiley (New York, 1962)

\bibitem{FF4}
M. Ferraris and M. Francaviglia,
Gen. Rel. Grav. {\bf 22}(9)\, 1990, p. 965

\bibitem{FFB}
M. Ferraris, M. Francaviglia and A. Borowiec (in preparation)

\bibitem{FK1}
M. Ferraris and J. Kijowski,
Gen. Rel. Grav. {\bf 14}(1)\, 1982, p. 37

\bibitem{FK2}
M. Ferraris and J. Kijowski,
Gen. Rel. Grav. {\bf 14}(2)\, 1982, p.165

\bibitem{Fer}
M. Ferraris,
{\it Affine Unified Theories of Gravitation and Electromagnetism}, Proceedings of
Journ\'ees Relativistes 1983, Ed. S. Benenti at al., Pitagora Editrice Bologna, 1985


\bibitem{Murphy}
G.L. Murphy, Int. Jour. of Theor. Phys. {\bf 29}(9)\, 1990, p.1003

\bibitem{Novotny}
J. Novotny, Int. Jour. of Theor. Phys. {\bf 32}(6)\, 1993, p. 1033

\end{thebibliography}
\end{document}